\documentclass{webofc}
\usepackage[varg]{txfonts}   
\usepackage{slashed}
\newcommand{\mcl}[1]{\mathcal{#1}}
\newcommand{\del}{\partial}
\newcommand{\n}{\notag \\}
\newcommand{\fsl}[1]{\slashed{#1}}

\newcommand{\la}{\ensuremath{\lambda}}

\newcommand{\ph}{\ensuremath{\phi}}

\newcommand{\ch}{\ensuremath{\chi}}
\newcommand{\ps}{\ensuremath{\psi}}

\begin{document}
\title{Atomki anomaly and the Secluded Dark Sector}
%
%

\author{
  \firstname{Yasuhiro} \lastname{Yamamoto}\fnsep\thanks{\email{yamayasu@yonsei.ac.kr}}
}

\institute{
  Department of Physics and IPAP, Yonsei University, Seoul 03722, Republic of Korea
}

\abstract{%
  The Atomiki anomaly can be interpreted as a new light vector boson.
	If such a new particle exists, it could be a mediator between the Standard Model sector and the dark sector including the dark matter.
	We discussed some simple effective models with these particles.
	In the models, the secluded dark matter models are good candidates to satisfy the thermal relic abundance.
	In particular, we found that the dark matter self-interaction can be large enough to solve the small scale structure puzzles if the dark matter is a fermion. 
}
\maketitle
\section{Introduction}

A nuclear experiment in Hungary has reported the observation of a new resonance in the decay of excited $^8$Be nuclei~\cite{Krasznahorkay:2015iga}.
This is called the Atomki anomaly named after the institute where it is observed.
The resonance can be simply interpreted as a new weakly interacting light particle of 16.7 MeV.

It is elucidated by Ref.~\cite{Feng:2016jff} that the new particle should be a vector boson.
They have also pointed out the coupling of the vector with proton is much smaller than that of neutron because of the strong constraint by the $\pi^0$ decay experiment~\cite{Batley:2015lha}.
Hence, the vector boson is also referred as the protophobic light vector boson.
The vector boson is not only the protophobic but also the leptophobic.
In particular, the coupling with the neutrinos should be smaller than about $10^{-5}$, instead, the coupling with the electron is O($10^{-4}$).
Because of this property, the model building is difficult; for the details, see also Refs.~\cite{Feng:2016ysn}.

If we suppose that the vector is the gauge boson of a broken U(1) symmetry, the Lagrangian can be written as
\begin{align}
 \mcl{L}_X =&
  -\frac{1}{4} X^{\mu\nu}X_{\mu\nu}
	+(D_\mu S)^\dag (D^\mu S) +\mu_S^2 |S|^2 -\frac{\la_S}{2} |S|^4 \n
 =&
   -\frac{1}{4} X^{\mu\nu}X_{\mu\nu} +\frac{m_X^2}{2} X^\mu X_\mu 
	 +\frac{1}{2} (\del_{\mu} s)^2 -\frac{m_s^2}{2} s^2
	 +g_X m_X s X^\mu X_\mu -\frac{g_X m_s^2}{2 m_X} s^3 +\cdots
\end{align}
where $X^{\mu\nu}$ is the field strength tensor of the new gauge boson $X$.
The scalar field $S$ is expanded as $S=(v_s + s)/\sqrt{2}$ in the unitarity gauge, and $D_\mu S=(\del_\mu +i g_X X_\mu) S$.
Some terms irrelevant in our computation are suppressed here.
The other parameters are defined as $v_s^2 = 2\mu_S^2 /\la_S$, $m_s^2 = \la_S v_s^2$, and $m_X = g_X v_s$.
We studied simple dark matter models including the above Lagrangian~\cite{Kitahara:2016zyb}.

If the dark matter directly interacts with $X$, it is difficult to satisfy the thermal relic scenario because the interaction strength of the new vector is comparable to the weak interaction, namely, $g_X/m_X = 1/O(100)$ GeV; see Refs.~\cite{Jia:2016uxs} for exceptions.
In the rest of this article, we discuss the models of the real scalar and the Majorana fermion dark matter.
Since they are neutral under the new gauge symmetry U(1)$_X$, the models also require other particles which mediate the interaction between the dark matter and $X$.
They can be easily compatible with the thermal relic scenario.
These models are called the secluded dark matter~\cite{Pospelov:2007mp}.
The given models are easily implemented in not only the explicit models of the protophobic light vector but also other models since the dark matter properties are mostly determined in the dark sector.

\section{Real scalar dark matter}

The Lagrangian of the real scalar dark matter model is 
\begin{align}
 \mcl{L}_{\phi} =&
  \frac{1}{2} (\del^\mu \ph)^2 -\frac{m_{\phi}^2}{2} \ph^2 -\frac{\la_{\ph S} m_X }{2 g_X} s\ph^2
 -\frac{\la_{\ph S}}{4} s^2 \ph^2
 -\frac{\la_\ph}{4!} \ph^4 +\mcl{L}_X,
\end{align}
where we impose a $Z_2$ symmetry ($\phi \leftrightarrow -\phi$) to stabilize the dark matter.
The coupling $\la_{\ph S}$ is given by the scalar four point interaction.
The dark Higgs boson $s$ play a role of the mediator.

\begin{figure}[ht]
\centering
  \includegraphics[scale=.52]{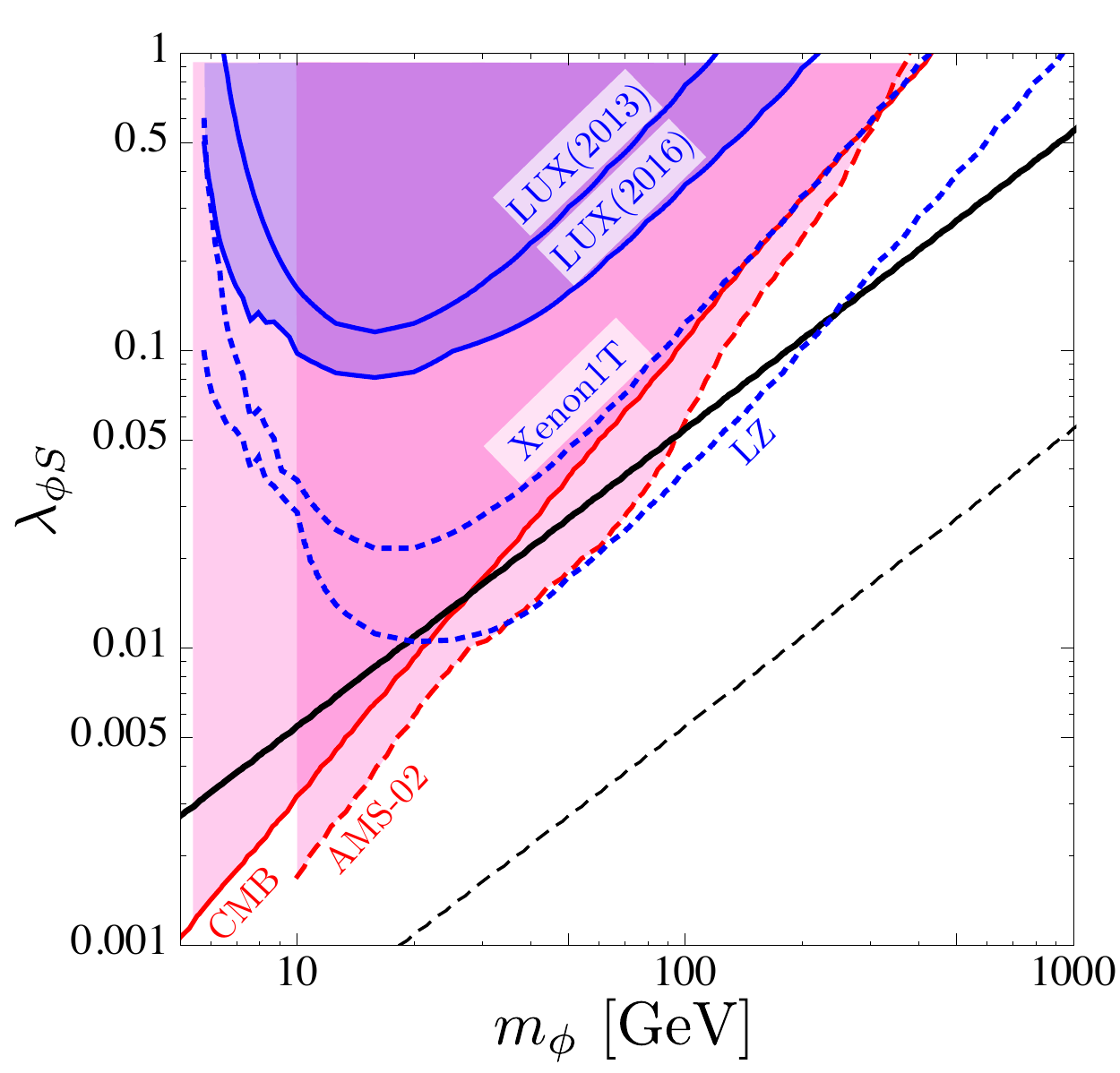} \qquad
  \includegraphics[scale=.5]{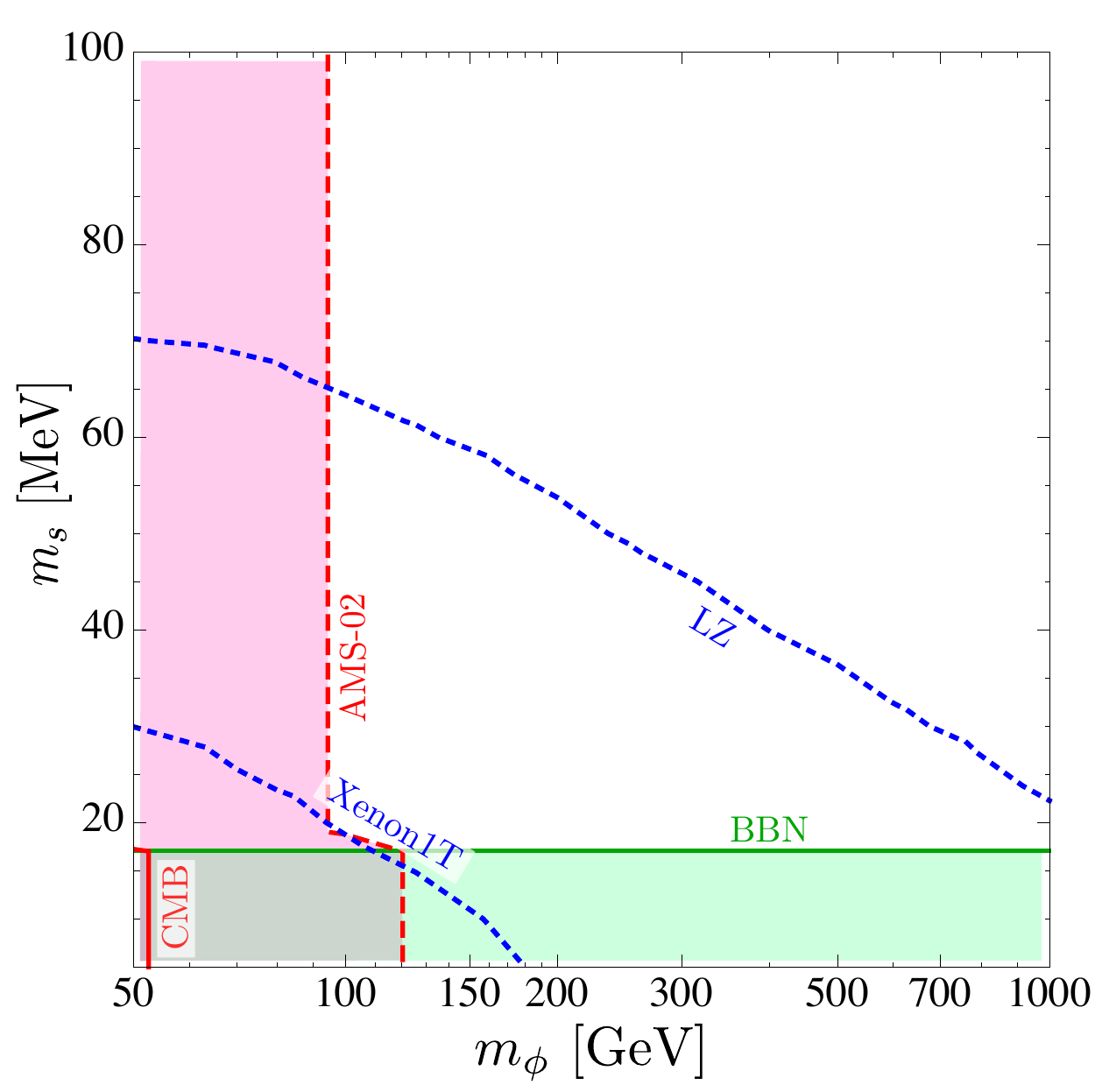}
\caption{
  Experimental bounds on the real scalar dark matter.
  The left figure is the bounds on the typical dark matter coupling $\la_{\ph S}$ as a function of the dark matter mass with $m_s=50$ MeV.
	The blue/red region is bounded by the current direct/indirect dark matter searches.
	The projected direct detection bounds by the XENON1T and the LZ experiments are shown with the blue dotted lines.
  On the solid black line, the dark matter satisfies the observed thermal relic abundance.
  On the dotted black line, the relic abundance is 2 orders of magnitude larger than the observed one.
  The dark Higgs mass dependence of the expected direct detection bounds is shown in the right figure.
  The coupling $\la_{\ph S}$ is chosen to satisfy the thermal relic abundance, i.e. on the black solid line of the left one.
  The blue dotted lines are the projected direct detection bounds.
	The red region is excluded by the indirect detections.
	In the green region, the lifetime of the dark Higgs is larger than 1 sec.
}
\label{FigReal}
\end{figure}

The annihilation of the dark matter is dominated by the s-wave processes: $\phi \phi \to XX$ and $\phi \phi \to  s s$.
Their final states are two and four $e^+ e^-$ pairs, respectively.
The direct detection is induced by the 1-loop diagram mediated by the $sXX$ vertex.
The experimental bounds on the models are shown in Fig.~\ref{FigReal}.

The left figure is the bounds for $m_s =50$ MeV.
The thermal relic abundance can be satisfied if the dark matter is heavier than about 100 GeV.
The indirect detection bounds with the $e^+ e^-$ spectrum by AMS-02 and Planck exclude the scenario up to there~\cite{Accardo:2014lma}.
The current LUX and the future XENON1T bounds are much weaker than the indirect bounds~\cite{Akerib:2016vxi}.
The current strongest bound by PandaX-II is the middle of them~\cite{RefPanda}.
However, the prospected LZ bound is stronger than them~\cite{Akerib:2015cja}.

The right figure shows us the $m_s$ dependence.
The coupling $\la_{ph S}$ is fixed to obtain the thermal relic abundance.
We find that the experimental reach of the dark matter mass strongly depends on the mass of the dark Higgs.

\section{Majorana fermion dark matter}

The Lagrangian of the Majorana fermion dark matter model is
\begin{align}
 \mcl{L}_{\chi} =&
   \bar{\ps} (i\fsl{\del} -g_X \fsl{X} -m_\ps )\ps
	+\frac{1}{2} \bar{\ch}(i\fsl{\del} -m_\ch)\ch 
	-y (S \bar{\ps} \ch +S^\dag \bar{\ch} \ps)  +  \mcl{L}_X,
\end{align}
where the Yukawa coupling $y$ can be chosen as real and positive without loss of generality.
The Dirac fermion $\ps$ is the additional mediator.
After the spontaneous symmetry breaking of U(1)$_X$, the model includes three Majorana fermions; the lightest one is the dark matter.

\begin{figure}[ht]
\centering
  \includegraphics[scale=.52]{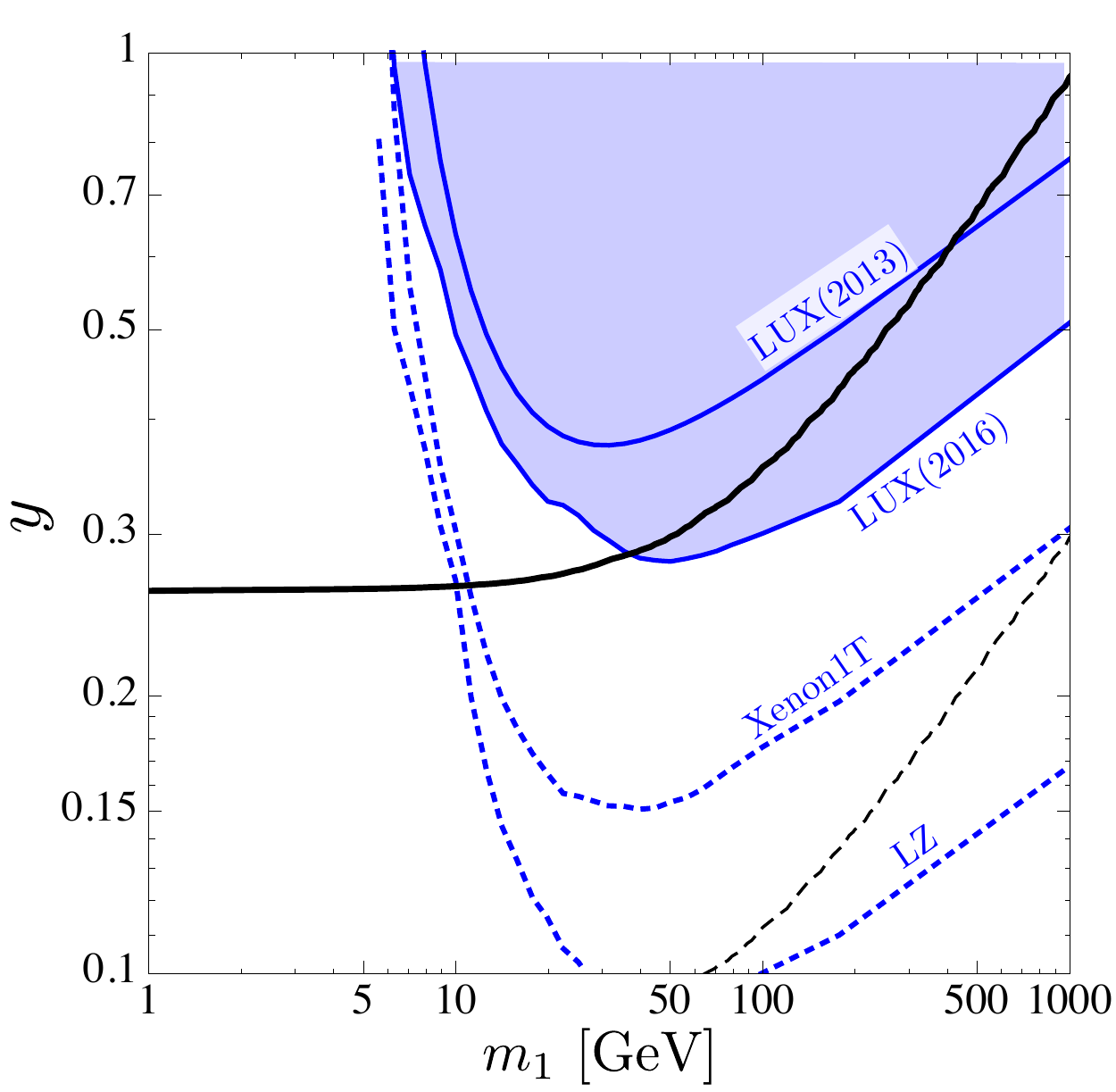} \qquad
  \includegraphics[scale=.5]{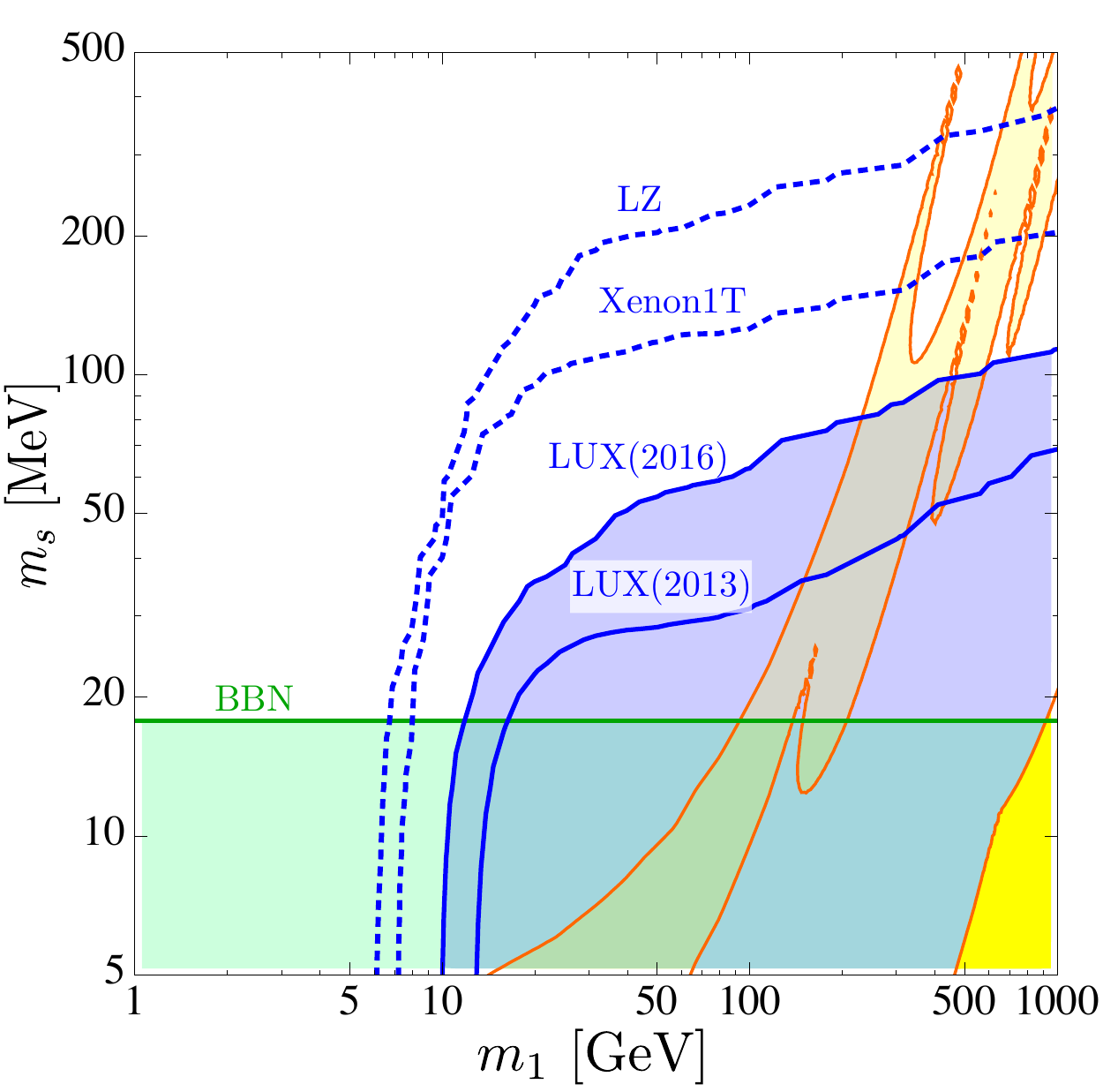}
\caption{
  Experimental bounds on Majorana fermion dark matter.
	The lines and regions are similar to Fig.~\ref{FigReal}.
	In the right figure, the small scale structure puzzles can be solved in two yellow regions at the middle and the corner, where $g_X = 10^{-3}$ and $10^{-2}$ respectively.
}
\label{FigMajo}
\end{figure}

The leading contribution of the annihilation is given by the p-wave, so that, the indirect bounds are negligibly weak.
The direct detection is given by the same diagram with the real scalar case.
However, as shown in Fig.~\ref{FigMajo}, the experimental bounds are significantly stronger than the previous one.

In these figures, the additional Majorana fermions are 100 GeV heavier than the dark matter.
If $m_s =50$ MeV, the left figure tell us that the dark matter should be lighter than a few tens GeV.
The upper bound becomes about 10 GeV in the near future.

The $m_s$ dependence is shown in the right figure.
Since the indirect detection bounds can be neglected, it is difficult to exclude the dark matter lighter than about 5 GeV.
If the dark matter is O(100) GeV or heavier, the dark Higgs should be heavier than about 100 MeV.
The bound reaches up to 200--400 MeV by the prospected LZ bound.

Instead, the dark matter annihilates through the p-wave, the dark matter self- scattering is the s-wave.
Besides, the cross section is largely enhanced by the Sommerfeld enhancement.
The large self- interaction of the dark matter is required to solve the small scale structure puzzles~\cite{Tulin:2013teo}.
If $g_X = 10^{-3}$, the self- interaction cross section becomes large enough to explain the structure of the dark matter of O(100) GeV.

\section{Summary}

\begin{figure}[ht]
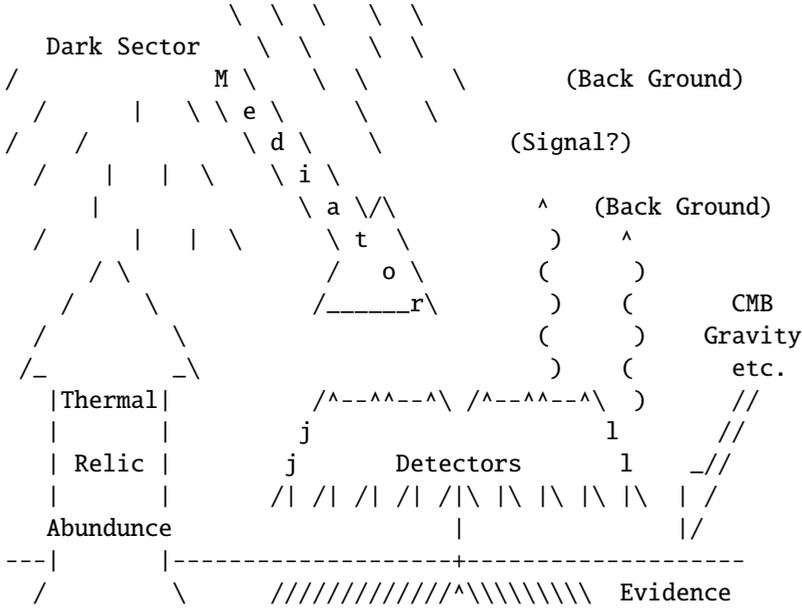

\centering
\begin{verbatim}
                           \  \  \   \  \
              Dark Sector    \  \    \  \ 
           /              M \    \  \      \       (Back Ground)
             /      |   \ \ e \     \    \            
           /    /           \ d \    \         (Signal?)
             /    |   |  \    \ i \           
                 |              \ a \/\          ^   (Back Ground)
             /      |   |  \      \ t  \          )    ^ 
                 / \              /   o \        (      )    
               /     \           /______r\        )    (       CMB
             /         \                         (      )    Gravity
            /_         _\                         )    (       etc.
              |Thermal|          /^--^^--^\ /^--^^--^\  )      //
              |       |         j                     l       //
              | Relic |        j       Detectors       l    _// 
              |       |       /| /| /| /| /|\ |\ |\ |\ |\  | /
              Abundunce                    |               |/
           ---|       |--------------------+--------------------
             /         \      /////////////^\\\\\\\\\  Evidence
\end{verbatim}
\caption{
  The schematic figure of the models discussed in this paper.
	The astrophysical observations clarify the dark components in our universe.
	The dark sector including the dark matter could be thermally produced.
	We suppose this sector interacts with the Standard Model via the weakly interacting light mediator, namely, the protophobic vector boson of 16.7 MeV.
	This figure is inspired by the sketch of photosynthesis in Ref.~\cite{RefMasaru}
}
\label{FigMasaru}
\end{figure}

A clear peak in the $e^+ e^-$ spectrum of excited $^8$Be has been observed by a Hungarian nuclear experimental group.
It looks not a statistical fluctuation like many tail events but a physical event.
The peak can originate from the weakly interacting light new vector boson.
Such a new particle is an attractive candidate of new physics.
As a fascinating scenario, we have studied the thermal relic dark matter models in which the vector boson is a mediator between the dark sector and the Standard Model, see Fig.~\ref{FigMasaru}.
We have discussed a real scalar and a Majorana fermion as the dark matter.
These models are similar each other.
However, they have clearly different behavior.

%

For the real scalar model, the thermal relic scenario is excluded by the indirect detections if the dark matter is lighter than about 100 GeV.
The direct detection bound can be stronger than the indirect bounds for the future LZ result.
The reach of the LZ bound is sensitive to the dark Higgs mass.

For the Majorana fermion model, the indirect detection bounds can be negligible because the annihilation is the p-wave.
The direct detection bounds exclude the dark matter of heavier than several tens GeV, while it is impossible to constrain the region where the dark matter is lighter than about 5 GeV.
In order to evade these experimental bounds for the heavier dark matter, we need to consider the dark Higgs of a few hundred MeV.
With the self-interaction, the thermal relic dark matter solves the small scale structure problem if its mass is several hundred GeV.



\begin{thebibliography}{99}
 
\bibitem{Krasznahorkay:2015iga}
  A.~J.~Krasznahorkay {\it et al.},
  Phys.\ Rev.\ Lett.\  {\bf 116} (2016) no.4,  042501
  doi:10.1103/PhysRevLett.116.042501
  [arXiv:1504.01527 [nucl-ex]];
  A.~J.~Krasznahorkay {\it et al.},
  EPJ Web Conf.\  {\bf 142} (2017) 01019.
  doi:10.1051/epjconf/201714201019 .

\bibitem{Feng:2016jff}
  J.~L.~Feng, B.~Fornal, I.~Galon, S.~Gardner, J.~Smolinsky, T.~M.~P.~Tait and P.~Tanedo,
  Phys.\ Rev.\ Lett.\  {\bf 117} (2016) no.7,  071803
  doi:10.1103/PhysRevLett.117.071803
  [arXiv:1604.07411 [hep-ph]].
 
\bibitem{Batley:2015lha}
  J.~R.~Batley {\it et al.} [NA48/2 Collaboration],
  Phys.\ Lett.\ B {\bf 746} (2015) 178
  doi:10.1016/j.physletb.2015.04.068
  [arXiv:1504.00607 [hep-ex]].

\bibitem{Feng:2016ysn}
  J.~L.~Feng, B.~Fornal, I.~Galon, S.~Gardner, J.~Smolinsky, T.~M.~P.~Tait and P.~Tanedo,
  Phys.\ Rev.\ D {\bf 95} (2017) no.3,  035017
  doi:10.1103/PhysRevD.95.035017
  [arXiv:1608.03591 [hep-ph]];
  B.~Fornal,
  Int.\ J.\ Mod.\ Phys.\ A {\bf 32} (2017) 1730020
  doi:10.1142/S0217751X17300204
  [arXiv:1707.09749 [hep-ph]].

\bibitem{Kitahara:2016zyb}
  T.~Kitahara and Y.~Yamamoto,
  Phys.\ Rev.\ D {\bf 95} (2017) no.1,  015008
  doi:10.1103/PhysRevD.95.015008
  [arXiv:1609.01605 [hep-ph]].

\bibitem{Jia:2016uxs}
  L.~B.~Jia and X.~Q.~Li,
  Eur.\ Phys.\ J.\ C {\bf 76} (2016) no.12,  706
  doi:10.1140/epjc/s10052-016-4561-3
  [arXiv:1608.05443 [hep-ph]];
  O.~Seto and T.~Shimomura,
  Phys.\ Rev.\ D {\bf 95} (2017) no.9,  095032
  doi:10.1103/PhysRevD.95.095032
  [arXiv:1610.08112 [hep-ph]].

\bibitem{Pospelov:2007mp}
  M.~Pospelov, A.~Ritz and M.~B.~Voloshin,
  Phys.\ Lett.\ B {\bf 662} (2008) 53
  doi:10.1016/j.physletb.2008.02.052
  [arXiv:0711.4866 [hep-ph]].

\bibitem{Accardo:2014lma}
  L.~Accardo {\it et al.} [AMS Collaboration],
  Phys.\ Rev.\ Lett.\  {\bf 113} (2014) 121101.
  doi:10.1103/PhysRevLett.113.121101;
  M.~Aguilar {\it et al.} [AMS Collaboration],
  Phys.\ Rev.\ Lett.\  {\bf 113} (2014) 121102.
  doi:10.1103/PhysRevLett.113.121102;
  P.~A.~R.~Ade {\it et al.} [Planck Collaboration],
  Astron.\ Astrophys.\  {\bf 594} (2016) A13
  doi:10.1051/0004-6361/201525830
  [arXiv:1502.01589 [astro-ph.CO]];
  M.~Kawasaki, K.~Nakayama and T.~Sekiguchi,
  Phys.\ Lett.\ B {\bf 756} (2016) 212
  doi:10.1016/j.physletb.2016.03.005
  [arXiv:1512.08015 [astro-ph.CO]].

\bibitem{Akerib:2016vxi}
  D.~S.~Akerib {\it et al.} [LUX Collaboration],
  Phys.\ Rev.\ Lett.\  {\bf 118} (2017) no.2,  021303
  doi:10.1103/PhysRevLett.118.021303
  [arXiv:1608.07648 [astro-ph.CO]];
  E.~Aprile {\it et al.} [XENON Collaboration],
  JCAP {\bf 1604} (2016) no.04,  027
  doi:10.1088/1475-7516/2016/04/027
  [arXiv:1512.07501 [physics.ins-det]].

\bibitem{RefPanda}
   X.~Cui {\it et al.} [PandaX-II Collaboration],
  arXiv:1708.06917 [astro-ph.CO].

\bibitem{Akerib:2015cja}
  D.~S.~Akerib {\it et al.} [LZ Collaboration],
  arXiv:1509.02910 [physics.ins-det].

%
\bibitem{Tulin:2013teo}
  S.~Tulin, H.~B.~Yu and K.~M.~Zurek,
  Phys.\ Rev.\ D {\bf 87} (2013) no.11,  115007
  doi:10.1103/PhysRevD.87.115007
  [arXiv:1302.3898 [hep-ph]];
  M.~Kaplinghat, S.~Tulin and H.~B.~Yu,
  Phys.\ Rev.\ Lett.\  {\bf 116} (2016) no.4,  041302
  doi:10.1103/PhysRevLett.116.041302
  [arXiv:1508.03339 [astro-ph.CO]].

\bibitem{RefMasaru}
  \url{https://en.wikipedia.org/wiki/Sexy_Commando_Gaiden:_Sugoi_yo!!_Masaru-san}
\end{thebibliography}
\end{document}